\documentclass{optica-article}

\journal{opticajournal}

\articletype{Research Article}

\usepackage{physics}
\usepackage{lineno}
\usepackage{float}
\usepackage{braket}
\usepackage{ulem}
\usepackage{todonotes}

\begin{document}

\title{Transient fluorescence with a single trapped ion}

\author{Nicol\'{a}s A. Nu\~{n}ez Barreto,\authormark{1,2,*} Lucas T. Giardino,\authormark{1} Carla J. Crucianelli,\authormark{1} Muriel Bonetto,\authormark{1,2} Mart\'{i}n Drechsler,\authormark{1,2} and Christian T. Schmiegelow\authormark{1,2}}


\address{\authormark{1} Universidad de Buenos Aires, Facultad de Ciencias Exactas y Naturales, Departamento de F\'{i}sica, Laboratorio de Iones y \'{A}tomos Fr\'{i}os, Pabell\'{o}n 1, Ciudad Universitaria, 1428 Buenos Aires, Argentina\\

\authormark{2} CONICET - Universidad de Buenos Aires, Instituto de F\'{i}sica de Buenos Aires (IFIBA), Pabell\'{o}n 1, Ciudad Universitaria, 1428 Buenos Aires, Argentina}

\email{\authormark{*}nnunez@df.uba.ar} 



\begin{abstract}
In this paper we present a method to measure transient fluorescent dynamics with single trapped ions in a Paul trap. We use $^{40}$Ca$^+$ ions which exhibit a $\Lambda$-type three-level system and measure the characteristic optical pumping times between the ground $S_{1/2}$ and the meta-stable $D_{3/2}$ levels. We prepare one of these states, then pump it to the opposite via the excited $P_{1/2}$ state. By measuring the scattered photons of the ion, we retrieve transient curves of the atomic fluorescence. These curves provide fundamental information about the atomic system, such as branching fractions and  excited-state lifetimes, as well as experimental parameters like the efficiency of the detection system and the saturation parameter of one of the transitions. Finally, we study the time dependent fluorescence as a function of optical power and detuning of the lasers and find a very good agreement with simulating the dynamics via a three-level open quantum system through a set of optical Bloch equations. 
Being able to record time dependent fluorescence is of particular interest as it contains information about the temperature, cooling and heating dynamics of the ion.
\end{abstract}

\section{Introduction}
Spurred by the invention of the laser,
the interaction of atomic systems with light sources has allowed for significant advances in the last decades, leading to the coherent control of single quantum systems~\cite{wineland2013nobel}.
Due to its high coherence times and, therefore, narrow bandwidths, laser spectroscopy has not only become one of the most reliable methods to study atomic structures and dynamics, but also a platform to explore fundamental physics using optical clocks~\cite{ludlow2015optical} as well as applications of quantum mechanics including quantum computing~\cite{cirac1995quantum,haffner2008quantum}, quantum simulation and quantum thermodynamics~\cite{blatt2012quantum,ramm2014energy}.

The absorption probability of photons by a single atom can exhibit varying time dynamics of the atomic populations, particularly when optical pumping plays an important role~\cite{wineland1979laser}. Typically, for dipole allowed transitions, these dynamics have time scales in the order of a few ns to $10~\mu$s. These specific time dynamics can be used as a tool to measure atomic properties such as the precise determination of the branching ratio of an excited state~\cite{gerritsma2008precision,ramm2013precision,likforman2016precision}, performing direct atomic spectroscopy~\cite{pruttivarasin2014direct}, measuring dipole matrix elements\cite{hettrich2015measurement}, estimating of the temperature of a single atom~\cite{RoffnagelCPTthermometry}, detecting trapped-ion states using the motion of the ion~\cite{hume2011trapped} or even performing direct measurements of the motion near the quantum level~\cite{cerchiari2021measuring}. 

In this work we use a single trapped calcium ion to study the time dependent atomic dynamics of a $\Lambda$-type system formed by the fundamental $4^2S_{1/2}$ state and the meta-stable $3^2D_{3/2}$ state which connect to the $4^2P_{1/2}$ excited state via dipolar transitions. We use a commercial FPGA-based ARTIQ control system~\cite{bourdeauducq2016artiq,kasprowicz2020artiq} to produce nanosecond timed digital sequences which control the laser power and detunings via acousto-optic modulators, while recording fluorescence by time stamping events detected on a photomultiplier tube. This allows us to measure and determine laser switching times and collection efficiencies as well as branching fractions and saturation parameters of the atomic transitions.  We also observe the dependence of the time dynamics on the power and detuning of the respective lasers and find a very good agreement with a three-level optical Bloch equations model. In the future we expect to use this system to study thermal and coherent motional dynamics of ionic crystals, as well as a tool to optimize atomic spectroscopy~\cite{barreto2022three}.

\section{Theory}

We describe the dynamics of the atomic populations by a set of differential equations for the elements of the density matrix $\rho$ of the system, namely, the \textit{optical Bloch equations} (OBE). We consider a three-level system consisting of two fundamental states $\ket{S}$ and $\ket{D}$ connected with the excited state $\ket{P}$ by two different lasers, one near 397~nm named \textit{Doppler}  and one  near 866~nm named \textit{repump} due to their physical roles in observing resonance fluorescence, as depicted in Fig. \ref{fig:fig1}(a). Here we do not take into account the magnetic degeneracy nor Zeeman splittings as the dynamics of interest are well described by a three-level system.

Writing the Hamiltonian of the system in a rotating frame of reference with the frequencies of the lasers, and using the rotating-wave approximation, one obtains 
\begin{equation}
    H/\hbar = \Delta_{\textrm{dop}}\ketbra{S}{S}+\Delta_{\textrm{rep}}\ketbra{D}{D} + \frac{\Omega_{\textrm{dop}}}{2}(\ketbra{S}{P}+\ketbra{P}{S})+\frac{\Omega_{\textrm{rep}}}{2}(\ketbra{D}{P}+\ketbra{P}{D}),
\end{equation}
where $\Omega_{\textrm{dop}}$, $\Omega_{\textrm{rep}}$ are the Rabi frequencies of both transitions (assumed real for simplicity), and $\Delta_{\textrm{dop}}$, $\Delta_{\textrm{rep}}$ are the detunings of the lasers, i.e., the difference between the frequency of the transition and that of the laser. The OBE arise from a master equation for $\rho$ which considers non-unitary dynamics as spontaneous emission and finite laser linewidths. This is done by adding a Linbladian superoperator to the unitary dynamics such that the final master equation is
\begin{equation}
    \frac{d\rho}{dt} = -\frac{i}{\hbar}\left[H,\rho\right]+\sum_m \left[ \mathcal{C}_m\rho\mathcal{C}_m^{\dagger} - \frac{1}{2}\left( \mathcal{C}_m^{\dagger}\mathcal{C}_m \rho + \rho \mathcal{C}_m^{\dagger} \mathcal{C}_m \right)  \right].
\end{equation}
The transition operators $\mathcal{C}_m$ model two non-unitary processes. First, the two spontaneous emission channels from the upper level to the lower levels. These have the form $\mathcal{C}_{\mathrm{SP}} = \sqrt{\Gamma_{\mathrm{SP}}}\ketbra{S}{P}$ and $\mathcal{C}_{\mathrm{DP}} = \sqrt{\Gamma_{\mathrm{DP}}}\ketbra{D}{P}$, where $\Gamma_{\mathrm{SP}}$ and $\Gamma_{\mathrm{DP}}$ are the linewidths of the respective transitions. Second, we consider finite laser-linewidths by adding broadenings to the $S$ and $D$ states through the operators $\mathcal{C}_{\textrm{dop}} = \sqrt{\Gamma_{\textrm{dop}}}\ketbra{S}{S}$ and $\mathcal{C}_{\textrm{rep}} = \sqrt{\Gamma_{\textrm{rep}}}\ketbra{D}{D}$, where $\Gamma_{\textrm{dop}}$ and $\Gamma_{\textrm{rep}}$ are the linewidths of the Doppler and the repump laser respectively.

We numerically integrate the OBE and compare the simulations with the experimental results. In the next section, we present the results of measured transitory dynamics and show how we can apply these methods to perform estimations of atomic parameters like optical pumping characteristic times and branching fractions, and experimental parameters such as detection efficiencies and saturation parameters.

\section{Experimental results}
\subsection{Experimental setup}
We performed experiments with single $^{40}$Ca$^+$ ions. These are trapped in an electrodynamical Paul trap which consists of an array of electrodes that provide a dynamical 3D confinement of the ions in its center. The trapping is accomplished by a radial confinement given by a ring-shaped electrode of $\sim 1.2$~mm in diameter carrying $\sim600~$V$_{\mathrm{pp}}$ at 22.1~MHz, and an axial confinement generated by two endcaps located $\sim 0.6$~mm from the center of the trap pointing in opposite directions, both with a DC voltage of $\sim1~$V. Also, there are additional DC electrodes near the trap to compensate for stray electrical fields and to position both the DC and the AC potential minima in the same location, thus avoiding excess micromotion and therefore, inefficient cooling. The trap design, discussed in~\cite{barreto2022three}, was inspired by the PTB Trap design discussed in~\cite{dolevzal2015analysis}.

\begin{figure}[h]
\makebox[\textwidth][c]{\includegraphics[width=1\columnwidth]{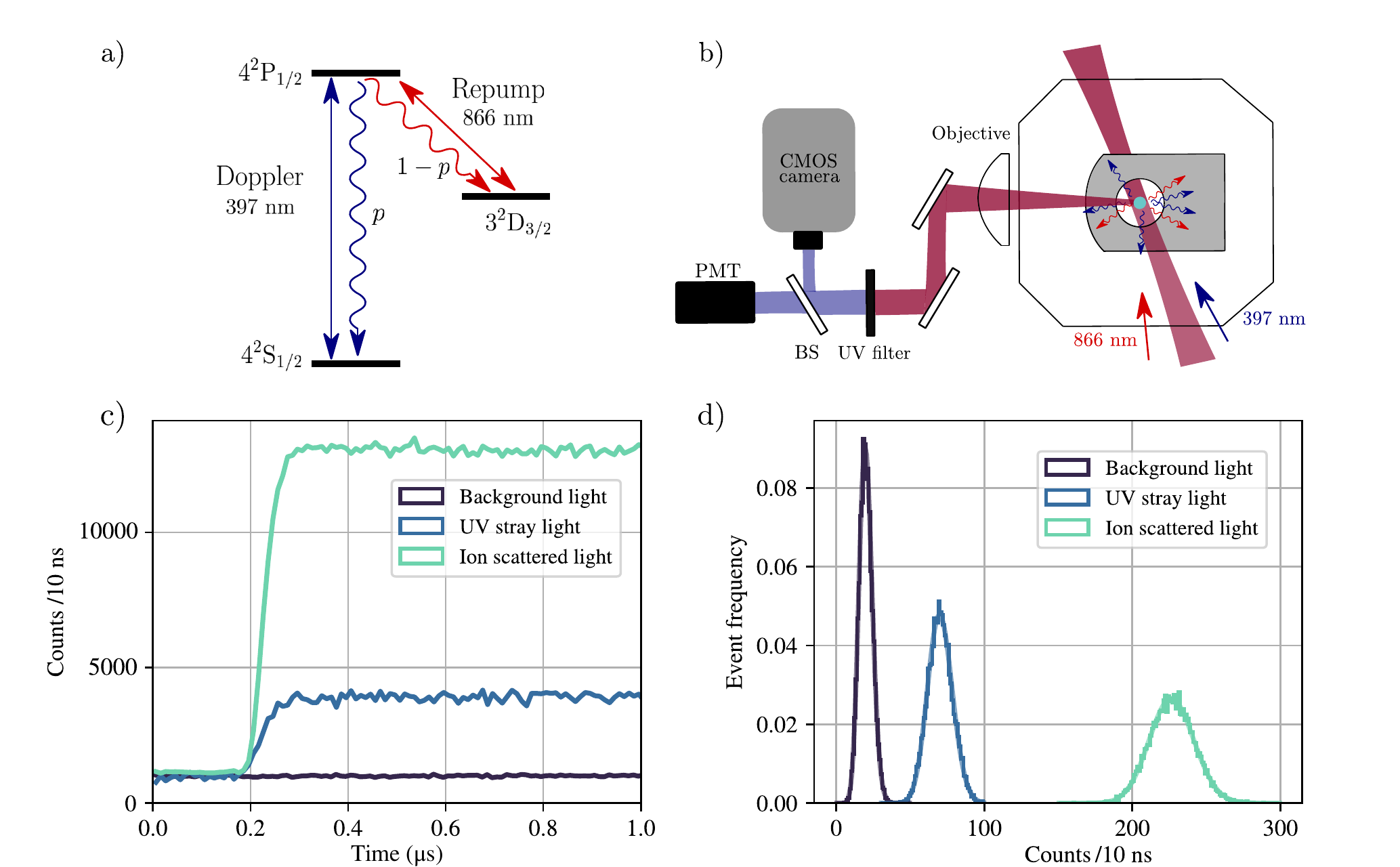}}
\caption{\label{fig:fig1}{(a) Relevant level scheme for $^{40}$Ca$^+$ ions. A $4^2S_{1/2}$ level is connected to a $4^2P_{1/2}$ level with a UV 397~nm (\textit{Doppler}) laser, from which it has a probability $p$ to decay to the same level and a probability $1-p$ to decay to a long-lived metastable $3^2D_{3/2}$ state. An IR 866~nm (\textit{Repump}) laser  depopulates the D state in order to have a closed fluorescence cycle. (b) Scheme of the experimental setup. The ion (light-blue) is trapped in the center of a ring-shaped Paul trap (gray), and illuminated with a pair of lasers in a collinear configuration. The ion scatters light and a part of it is collected with an objective lens, filtered with a 397~nm-interference UV filter and focused onto a photomultiplier tube (PMT) and a CMOS camera. (c) Time dependent collected light of a single fluorescent ion (aquamarine), the UV stray light (blue) and the background light (purple). We performed 50~M $1~\mu$s-long measurements to gain contrast in the transient part of the curves. (d) Histograms of detected counts in the stationary regime, with the respective Poisson distributions, for the same three cases, taken from a total of 3~M, $10~\mu$s-long measurements.}}
\end{figure}

In Fig. \ref{fig:fig1}(a) we show the relevant $\Lambda$-type three-level scheme for the system. The $4^2S_{1/2}$ ground state is dipole-connected with the excited $4^2P_{1/2}$ state by an ultraviolet (UV) 397~nm-laser. This excited state has a probability $p$ of decaying to the $S$ state, and a probability $1-p$ of decaying to the meta-stable long-lived $3^2D_{3/2}$ state. To avoid population to be pumped to the $D$ state, we use an infrared (IR) 866~nm-laser to repump the atomic population to the $P$ state, hence describing a closed fluorescence cycle. Other nearby-lying atomic levels do not play a role in these experiments as they are either far detuned or transitions to them are prohibited by selection rules.

Both lasers are frequency locked to two independent mid-finesse ($\sim 200$) Fabry-Perot cavities through Pound-Drevel-Hall method~\cite{drever1983laser} using Red Pitaya boards~\cite{redpitaya} adapted for this purpose \cite{luda2019compact}. This also reduces the linewidth of the lasers to $\sim100~$kHz. The beams go through two independent acousto-optic modulators (AOMs) in double-pass configuration which serve as fast optical switches, providing a high level of control in turning on and off the lasers, as well as a mechanism to change or sweep their frequencies a few tens of MHz.

In Fig 1(b) we show a sketch of the experimental setup. The ion is trapped in the center of the ring-trap and illuminated with the two lasers from one side. As it scatters part of the incoming light, UV and IR photons are emitted in all directions by spontaneous emission, a fraction of which are collected by an objective lens with a 50~mm focal length. The light is then filtered by a 397~nm-interference filter to detect only photons from the $S$-$P$ transition and is then split with a non-polarizing beamsplitter which transmits $\sim90$\% of the light to a photomultiplier tube (PMT) for fast time-tagged measurements, and reflects the remaining $\sim10$\% to a CMOS Andor camera for ion visualization and monitoring.

To perform the experiments we use an ARTIQ control system~\cite{bourdeauducq2016artiq,kasprowicz2020artiq}. It has two radio-frequency (RF) outputs which control the AOMs and one TTL input which is used to read the PMT. The system allows one to program and synchronize the tuning and switching of the AOMs and to time-tag the incoming photons.

\subsection{Steady-state fluorescence and laser switching timing}

Here we describe a basic characterization of the detection system, where we measure the background counts with the lasers off and compare it with the transitory of stray light from the UV laser and the fluorescence light of a single ion.

In Fig. \ref{fig:fig1}(c) we show the transient curves of the three cases. 
In all cases at $t=0$ we turn on different lasers and begin acquisition with the PMT for up to  1~$\mu$s. This sequence is repeated 50~M times to reduce statistical uncertainties and gain contrast in the transient zone. 
In purple, we show the background counts acquired by leaving all the lasers off. We obtain a measurement with a constant mean value as expected, which is due to spurious light and dark counts. In blue, we show the measurement of the UV stray light. 
At $t = 0$, the RF output of the AOM is turned on. We see that there is an initial dead time of $\sim 0.2$~$\mu$s and a rise time of $\sim 70$~ns. The dead time is mainly composed of an initial delay response time of the RF source ($\sim 50$~ns), plus delays due to propagation in coaxial cables ($\leq 10$~ns) and finally the propagation time it takes the acoustic wave to reach the beam in the crystal ($\sim 140$~ns). The rise time is mainly due to turn on behaviour of the RF source ($\sim 50$~ns) with a smaller contribution from the time that it takes the acoustic wave to go through the whole laser beam ($\sim 20$~ns). The two times associated with the AOM depend on the speed of sound in the material, the location of the beam within the crystal and the diameter of the beam.
Finally, in aquamarine, we show a transient curve of a fluorescent single ion. This is achieved by, first, turning on a 5~$\mu$s long IR pulse to prepare the S state, and, at $t=0$, turning on both the UV and IR lasers thereby creating a closed fluorescence cycle. We observe a transient time of $\sim$100~ns, which is a combination of the AOM rise time and the transient time of the atomic populations to reach a steady state according to the previously described optical Bloch equations. After that, the detected fluorescence, due mostly to the ion, has a constant mean value

Taking into account only the stationary values one can study the photon number statistics. In Fig. \ref{fig:fig1}(d) we show count frequency histograms obtained with 3~M measurements, each 10 $\mu$s long for the three light conditions. We use a binning window of 10~ns and plot the measurements along with theoretical Poisson distributions. We find the mean value of the measurements for each case, which match the expectation values of the distributions, to be \{19.54(1),69.76(2),226.24(1)\} counts per 10~ns for the background, stray and fluorescence lighting conditions respectively. As predicted by the nature of the uncorrelated events, the measurements are in good agreement with a Poisson distribution.

\begin{figure}[h]
\makebox[\textwidth][c]{\includegraphics[width=1\textwidth]{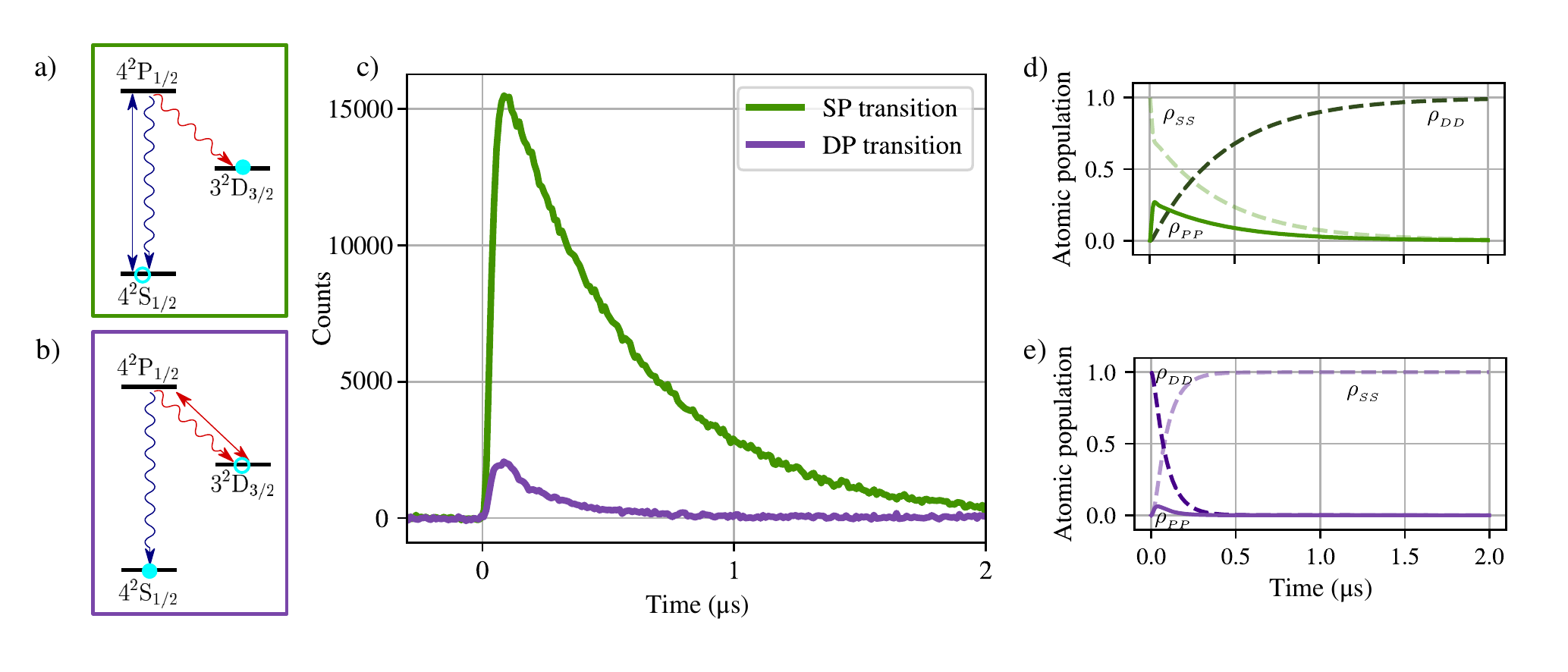}}
\caption{\label{fig:fig2}{Measurement of one $S$-$P$ and one $D$-$P$ transition. (a) Scheme to measure the $S$-$P$ transition. The ion is initialized in the $S$ state after scattering some UV photons. After that, the UV laser is turned on, and we start recording the scattered UV photons. Finally, the population ends in the $D$ state. (b) Scheme to measure the $D$-$P$ transition. The ion is instead initialized in the $D$ state and after that, the IR laser is turned on. At the same time, we begin recording the scattered UV photons. The ion scatters only one UV photon and the atom goes to the $S$ state. (c) Fluorescence as a function of time for the two optical pumping transitions. We repeat both measurements 50 M times and record the arrival of photons by time-tagging events. The measurements of the $S$-$P$ transition curve (green) and the $D$-$P$ transition curve (purple) are binned with a window of 10~ns. The areas are visibly different and this gives information on the value of $p$, i.e., the branching fraction of the atom. (d,e) Simulations of the atomic populations for the $S$-$P$ and $D$-$P$ optical pumping processes respectively. The fluorescence curves observed in (c) are proportional to the atomic populations of the excited state $\rho_{_{PP}}$ in each case. We also show the population of the other two levels, where one can appreciate optical pumping occurring.}}
\end{figure}

\subsection{Detection efficiency and branching ratio measurement via transient fluorescence}
To perform transitory experiments where optical pumping takes place, we initialize the ion either in the $S$ or the $D$ state with a strong 5~$\mu$s-long pulse of the 866 nm or the 397 nm laser respectively (pump laser), as seen in Fig. \ref{fig:fig2}(a) and Fig. \ref{fig:fig2}(b). This way, the atomic population is optically pumped to the desired initial state. After this, we proceed to turn on the other laser (probe laser) and at the same time, enable the detection of the PMT. The ion absorbs photons and the atomic population gets optically pumped to the opposite state. In both measurements we detect and timestamp only the UV scattered photons of the $S$-$P$ decay, and bin the data with a window of 10~ns. We call the first one "$S$-$P$ transition" and the second one "$D$-$P$ transition" since those are the transitions driven by the probe laser in each one.

To extract useful physical data from the transient phenomena, a background subtraction needs to be done. We split the background into two main contributions: the general background given by dark counts and light sources at the laboratory, and the scattered spurious UV light. The main difference between them is that the first one is constant through all the experiments and the second one can have a time-dependent behaviour, as seen previously, given by the slew in the AOM turn on.

For the $S$-$P$ experiments, we need to measure the AOM transient curves without the ion and use them to subtract it from the measurements. On the other hand, for the $D$-$P$, since the detection system filtrates the IR light, the background curve for this case will only be the general background, so it will be sufficient to subtract a constant value from the measurements.

In Fig. \ref{fig:fig2}(c) we see one $D$-$P$ curve in green and one $S$-$P$ curve in purple, both with the respective backgrounds subtracted. The measurement algorithm used is the following: first we turn on a strong 5~$\mu$s-long pump UV (for the $D$-$P$ curve) or IR (for the $S$-$P$ curve) pulse; then, we turn off the pump laser and wait for 2~$\mu$s to account for the transient turn-off response. Finally, we turn on the probe laser (IR for the $D$-$P$ curve and UV for the $S$-$P$ curve) and acquire the photon count on the PMT for 5~$\mu$s. We repeat this 50~M times to reduce statistical uncertainties. Also, once every 50 repetitions we turn on both lasers for 1~ms to avoid excess heating of the ion. The whole measurement for each curve takes approximately 1 hour and 20 minutes. In this time, the drift in frequency is below 1~MHz due to the Pound-Drever-Hall stabilization to vacuum-sealed cavities built from ultra low expansion glass. On the other hand, the drift in optical power is below 1\%, so both effects are ignored for the present analysis.

As seen in Fig. \ref{fig:fig2}(c), the area of both the $S$-$P$ and $D$-$P$ curves are noticeably different. This is well modeled by the simulations of the OBE shown in Fig. \ref{fig:fig2}(d) and \ref{fig:fig2}(e), where optical pumping manifests in the behavior of the population of the states. Following Ramm et.al.~\cite{ramm2013precision} we obtain the \textit{branching fraction} of the $P$ state, which represents the probability of the ion to decay to the $S$ level, by comparing the total counts detected in each transition, named $N_{\mathrm{SP}}$ and $N_{\mathrm{DP}}$ respectively. This way, one can calculate $p=N_{\mathrm{SP}}/(N_{\mathrm{SP}}+N_{\mathrm{DP}})$ which in our experiment yields $p=0.9357(14)$, in good agreement with previous measurements by Ramm et al. Here we only considered the uncertainty given by the shot noise of the measured counts, since, due to the short duration of the experiment, this is much greater than other sources of uncertainty like frequency or power drifts, electronics dead times, etc.


Furthermore, with the measurement of the $D$-$P$ transition alone we can estimate the efficiency of the detection system. This is a magnitude that is determined by the solid angle of the collection lens as well as the optical transparencies of the vacuum chamber windows, the beam-splitter and the PMT quantum efficiency. We directly measure this integrated efficiency by making use that if the ion is initially in the $D$ state, and the repump laser is turned on, it will only emit one UV photon before it is pumped to the $S$ state and goes dark. This way, by performing $N$ cycles, the ratio of photons detected over the number of cycles will represent the detection efficiency of our system. A measurement of  50 M cycles reveals a detection efficiency of 0.140(1)\%.

\begin{figure}[h]
\makebox[\textwidth][c]{\includegraphics[width=1\textwidth]{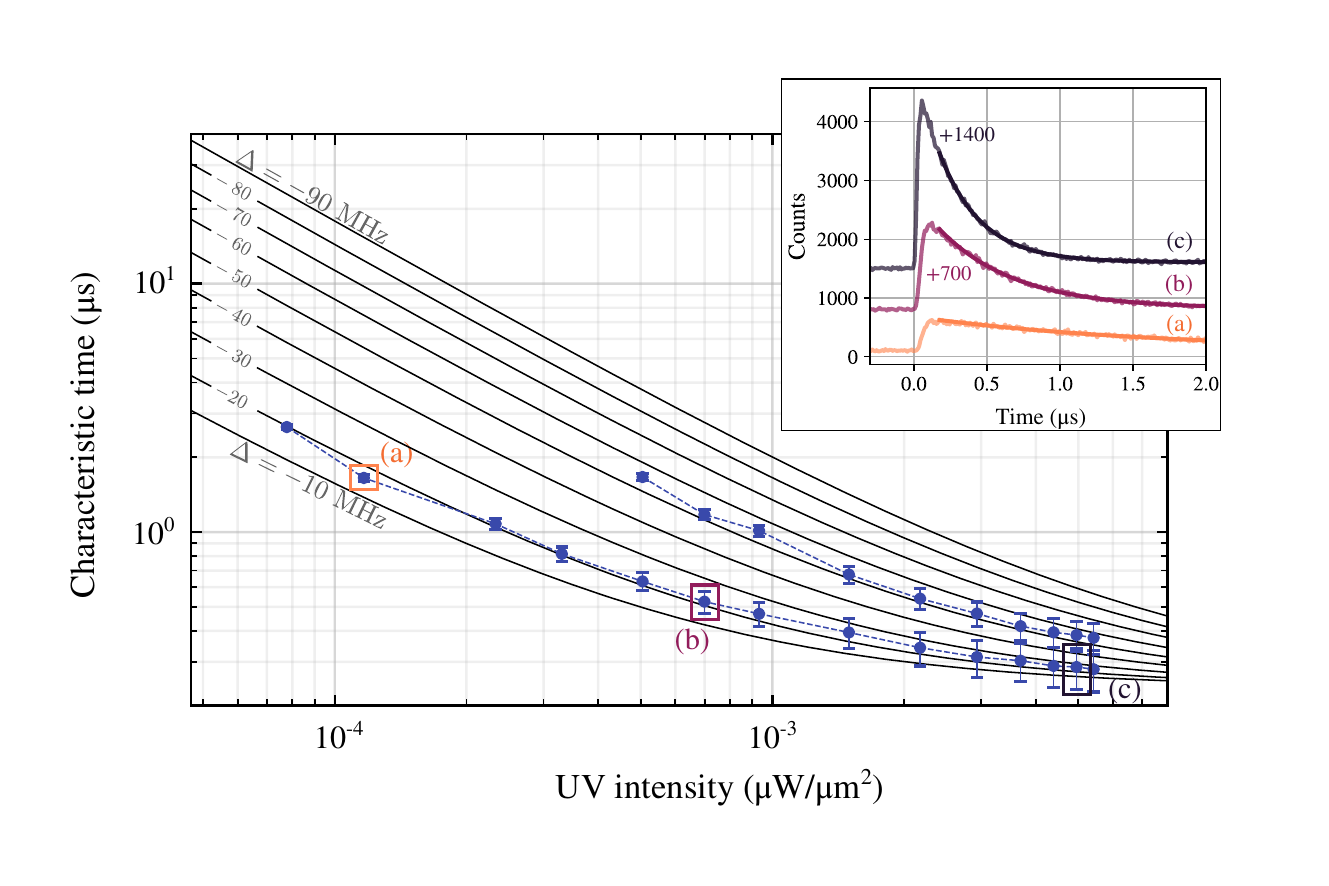}}
\caption{\label{fig:fig3}{Characteristic decay time of the fluorescence of a single ion varying the optical intensity (i.e. the Rabi frequency) of the UV laser driving the $S$-$P$ transition. We perform measurements for two different detunings, $\sim-20$~MHz and $\sim-50$~MHz. We also plot theoretical level curves for nine different detunings (black solid lines) obtained with simulations of the OBE. We find the measurements in good agreement with the simulations. In an inset we show curves for three different UV optical powers, displaced a certain amount in the $y$ axis for better appreciation as noted in each curve. The $x$ axis was also displaced such that $t=0$ corresponds to the beginning of the curve. The respective characteristic times retrieved from the curves are marked in the plot with the same colors.}}
\end{figure}

\subsection{Power and detuning behavior of transient fluorescence}

We now analyze the effect of laser power and detuning to the transient fluorescence behavior during optical pumping. We focus particularly on the $S$-$P$ transition.

The total number of scattered photons in each transition is independent of the power of the lasers, since the branching fraction depends only on the atom. However, if we vary the power of the laser, the characteristic time of the optical pumping varies. This is reflected in the characteristic time of the exponential decay of the atomic fluorescence. Using the $S$-$P$ optical pumping transition, we varied the optical power of the UV laser, and measured different curves. We fitted the last part of each curve with an exponential function $N(t)=N_0e^{-t/\tau}+C$, where $\tau$ represents the characteristic time of the optical pumping process and $N_0$ and $C$ are free fitting parameters. In the inset of Fig. \ref{fig:fig3} we show three different transitory curves of the $S$-$P$ transitions for three different optical powers of the UV laser. As can be seen, the characteristic time shortens with higher powers. In Fig. \ref{fig:fig3} we plot $\tau$ vs. the optical intensity of the laser, obtained by dividing the power with the beam area, and compare it to simulations to the optical Bloch equations. We measure the intensity dependence for two different detunings, finding the measurements in good agreement with the simulations corresponding to detunings of $\sim-20$~MHz and $\sim-50$~MHz.

Now, we show how one can use these transitory curves to measure laser parameters. We can get a comparison of the branching fraction from the $P$ level to the two lower levels by comparing the areas of the $S$-$P$ and $D$-$P$ curves of Fig. \ref{fig:fig2}(c). We determine the branching fraction from the $P$ level to the $S$ level to be $\sim14$ times greater than the one from the $P$ to the $D$. Therefore, we can approximate the dynamics of the $S$-$P$ transition as a fast two-level system dynamics between the $S$ and the $P$ state reaching a steady-state, with a slow decay channel to the $D$ state. This means that after one prepares the $S$ state and turns on the UV laser, one can approximate a stationary value for the atomic populations using the usual two-level system results. 
Measuring the characteristic decay time to the $D$ state, one can then directly measure the saturation parameter of the transition $S$-$P$, defined as $s={2\Omega^2}/({4\Delta^2+\Gamma^2})$, where $\Omega$ is the Rabi frequency of the transition of interest (proportional to the square root of the optical power), $\Delta$ is the detuning and $\Gamma$ is the corresponding linewidth. 
Then, the stationary value of the population of the excited state $\rho_{_{\mathrm{PP}}}$ after the UV laser is turned on can be expressed in terms of $s$ as $\rho_{_{\mathrm{PP}}}={s}/{2(1+s)}$. 
If we now consider the decay from that steady-state to the $D$ state through spontaneous emission at a rate $\Gamma_{\mathrm{DP}}=2\pi\times1.482$~MHz~\cite{hettrich2015measurement}, the measured characteristic decay time $\tau_{\mathrm{meas}}$ can be related with $\rho_{_{\mathrm{PP}}}$ as
$\tau_{\mathrm{meas}}^{-1}\approx\rho_{_{\mathrm{PP}}}(s)\Gamma_{\mathrm{DP}}$. 
This way, by measuring $\tau_{\mathrm{meas}}$ we can retrieve the saturation parameter $s$ without the need of estimating the detuning or the Rabi frequency. For instance, for the maximum intensity set in Fig. \ref{fig:fig3}, we obtain $\rho_{_{\mathrm{PP}}}=0.46(1)$, which yields a saturation parameter of $S=11.4(6)$. This value can be compared to an estimation using the saturation intensity of the transition. To calculate it, we take the transition linewidth  $\Gamma_{\mathrm{SP}}=2\pi\times 21.57(2)$~\cite{hettrich2015measurement} and the corresponding transition frequency $\omega_{\mathrm{SP}}=2\pi\times 755.222~$THz, and compute $I_{\mathrm{sat}}={\hbar \Gamma_{\mathrm{SP}} \omega_{\mathrm{SP}}^3}/({12\pi c^2})$, where $c$ is the speed of light, to obtain $I_{\mathrm{sat}}=45.1\times 10^{-5}$~$\mu$W/($\mu$m)$^2$. Then, taking the experimental intensity for that same data point  $I=5.42(12)\times 10^{-3}$~$\mu$W/($\mu$m)$^2$, we obtain a saturation parameter $S=I/I_{\mathrm{sat}}$ of 12.0(3), which is in good agreement with the value estimated using the characteristic decay time.

\section{Conclusions}
In this work we presented an experimental apparatus capable of  performing time dependent fluorescence measurements with transitory atomic dynamics of a single trapped ion. The dynamics are well modeled by three-level optical Bloch equations. This allows us to measure characteristic transient times of optical pumping dynamics which are in the order of a few ns to 10~$\mu$s.
These measurements allow us to estimate atomic properties like the branching fraction of excited states as well as directly measuring experimental parameters like the efficiency of the detection system, which is normally estimated using the single efficiencies of all optical elements.
Finally, we investigate the dependence of the transient times with the optical power and the detuning of the probe laser, and find it in good agreement with the simulations of the OBE.


\begin{backmatter}
\bmsection{Funding}
This work was supported by Agencia I+D+i Grants No. PICT 2018 - 3350 and No. PICT 2019 - 4349, Secretar\'{i}a de Ciencia y T\'{e}cnica, Universidad de Buenos Aires Grant No. UBACyT 2018 (20020170100616BA), and CONICET (Argentina). 

\bmsection{Acknowledgments}
We thank F. Schmidt-Kaler for his support and generosity as well as J. P. Paz, M. A. Latoronda, and A. J. Roncaglia for their unconditional help and backup in the setting up of the laboratory.

\bmsection{Disclosures}
The authors declare no conflict of interest.

\bmsection{Data availability} Data underlying the results presented in this paper are not publicly available at this time but may be obtained from the authors upon reasonable request.

\end{backmatter}

\bibliography{sample}

\end{document}